# Agent-based simulation of the learning dissemination on a Project-Based Learning context considering the human aspects

Laio O. Seman, Romeu Hausmann, Eduardo A. Bezerra

*Abstract*— This work presents an agent-based simulation (ABS) of the active learning process in an Electrical Engineering course. In order to generate input data to the simulation, an active learning methodology developed especially for part-time degree courses, called Project-Based Learning Agile (PBL$^A$), has been proposed and implemented at the Regional University of Blumenau (FURB), Brazil. Through the analysis of survey responses obtained over five consecutive semesters, using partial least squares path modeling (PLS-PM), it was possible to generate data parameters to use as an input in a hybrid kind of agent-based simulation known as PLS agent. The simulation of the scenario suggests that the learning occur faster when the student has higher levels of humanist's aspects as self-esteem, self-realization and cooperation.

## I. INTRODUCTION

In a reflection on the contextualization of engineers in the 21st century, [1] argues that engineers of tomorrow, and even today's engineers, will have to face deep and new challenges. For the author, the new engineers will have to deal with the stress of each day competing in a world of accelerating changes where they will have to solve problems without precedence of scope and scale. The recent and fast technological changes have resulted in transformations in several fields, and a new paradigm has driven the world to the information age [2], [3].

Concepts related to the humanization of engineering became important in preparing engineers for the new challenges they will face in the forthcoming decades. In a survey done with experts, the Great Engineering Challenges were defined for the coming years, presented without order of importance: storing solar energy, providing fusion energy, developing of carbon sequestration methods, managing the nitrogen cycle, providing access to clean water, developing better medicines, advancing in health informatics, safe cyberspace, preventing nuclear terror, restoring and improving urban infrastructure, reverse engineering the brain, improving virtual reality, advanced personalized learning, scientific discovery [1].

It is perceived that the challenges involve energy and sustainability, health care, and advances in the human capacity for self-knowledge, all of them interdisciplinary aspects [1].

Interdisciplinary is an important factor in building the necessary advances for new engineers, and it is necessary to build relationships with the natural, social, behavioral, computational and mathematical sciences. It is necessary for engineering as a whole to communicate with the other areas of knowledge to provide context for the students, providing the engineer with a view of their role in history and in the future [4].

To this ability to move beyond basic knowledge and achieve a level of understanding that allows the engineer to deal with new problems in an innovative and creative way, [5] calls "adaptive knowledge," and implies that such a factor must be the new goal to be achieved by educators in engineering, considering the human aspects of the education.

According to [6] the community has made significant advances in conducting studies related to engineering education and proposing goals that include this "adaptive knowledge", however, it has been less efficient at figuring out how to achieve those goals. In the view of [5], three educational components must be not only developed, but aligned to complement each other to achieve these objectives: curriculum, instruction and assessment.

In this context, this article proposes the focus on instruction, and how it is affected by individual (self-esteem and self-realization) and social (cooperation) student's aspects.

For this purpose, an active methodology called Project-Based Learning Agile was used as a case study over 5 semesters to model the students' relation to humanization using partial least squares path modeling (PLS-PM), later used as input data in an agent-based simulation, a social simulation that allows to evaluate how individuals act and interact with each other [7].

The objective of the social simulation raised in this work is to perceive the differences in the speed of learning of a group of agents through different levels of humanization and instruction quality level.

The remaining of the paper is organized as follows. Section II presents the background and characterization of the

Laio O. Seman is with the Department of Electrical Engineering, Federal University of Santa Catarina, Florianópolis, Brazil.

Romeu Hausmann is with the Department of Electrical Engineering of Regional University of Blumenau, Blumenau, Brazil.

Eduardo A. Bezerra is with the Department of Electrical Engineering, UFSC, Brazil, and with LIRMM, Université de Montpellier, France.

methodology, Section III presents the results of the survey data analysis, Section IV presents the agent-based simulation and Section V presents the conclusions.

## II. Background and Characterization

The pilot project took place in the Department of Electrical Engineering and Telecommunications of the Regional University of Blumenau - Brazil (FURB) during the first half of 2014 and had the involvement of the following courses: Power Electronics and Control and Servomechanisms. Since the initial implementation, the project was also applied in 2015/1, 2015/2, 2016/1 and 2016/2.

During the applications of the project the students were challenged with aspects related, but not limited, to designing a CC-CC converter and it's closed loop control and projecting a photovoltaic battery charger.

Two main aspects were considered important in the implementation of the projects: team's formation (integration among students) and differentiation (the same project for all teams, but with different requirements to allow experience exchanging).

Students who were not in the two courses intersection set could be divided into two groups: one group of those who already attended one course and are attending the other one; and the second group would be those who are attending one course and would attend the other in the future. Only the second could create a problem in the project's process. So there was a recommendation that the group should have members of both courses to share experiences and knowledge. The idea was to mitigated the issue.

### A. Project-Based Learning Agile

The part time profile of the Electrical Engineering major of FURB led to the creation of a new application model of Project-Based Learning. This, could be adapted to the student's profile, being dynamic and adaptive. Thus, its development was based on the principles of Agile practices.

The Agile manifesto [8] was used as a basis, which was adapted to better meet the expected requirements of such project, creating a methodology called Project-Based Learning Agile (PBL[A]), according to the following principles:

> **Individuals and interactions** over processes and tools
> **Working simulation** over comprehensive documentation
> **Student collaboration** over deadlines negotiation
> **Responding to change** over following a plan

In the adapted manifesto (as in the original), even if there is value in the items on the right hand side (those not in bold), the highest value is given to items on the left hand side (bold ones).

The Agile manifesto principles are important to give the project a flexible content. It is important that a project that can quickly adjust along the way, to meet potential difficulties that arise during the progress of the courses that form the project.

The difference of this proposal is its goal. It does not aim to make the students to follow the Agile methodology within the PBL, but to use the principles of the manifesto for the creation of PBL. As a result, only the instructors have contact with Agile, making the methodology transparent to students.

## III. PLS-PM as Parameter Modeling Tool

In order to assess the impact of the methodology for posterior use with the agent-based simulation, a survey first proposed in [14] was applied in the post-implementation phase of the projects. The idea is to measure the acceptance of PBL methodology and to identify possible skills developed by the students during the project.

As a way to operationalize the survey for the purpose of this work - to identify the elements that construct active learning in Electrical Engineering based on humanistic concepts - five dimensions were taken from the literature:

- PBL (composed by questions of prefix $P_x$)
- Learning (composed by questions of prefix $C_x$)
- Cooperation (composed by questions of prefix $G_x$)
- Self-esteem (composed by questions of prefix $E_x$)
- Self-realization (composed by questions of prefix $R_x$)

With the end of the project, the data analysis of the survey was made so that it was possible to assess the project's impact on the students. After five applications of PBL[A], 162 students' responses were collected.

Table 1 presents the used questionnaire and also the total percentage of agreement and disagreement for each question (including the five applications), ranging from 1 - "I completely disagree" to 5 - "I agree completely" to the first group and 1 - "Not satisfactory" to 5 - "Very satisfactory" for the remaining groups.

### A. Model Hypotheses

During the model conception, the classical education literature have been considered, and hypotheses have been raised to understand the importance of the human aspect in the educational process.

In order to measure the impact of the humanization on the proposed methodology and, consequently, on learning, the relevant dimensions of human relations (self-esteem, self-realization and cooperation) were grouped into a new dimension called "Humanization".

The new created dimension is based on the literature regarding the humanization of engineering education and the 21st century required skills [9]–[12]. Its creation considers intrapersonal and interpersonal skills, and seeks to understand how both influence student training.

TABLE I
PROCESS ASSESSMENT SURVEY

| Manifests | | Frequency (%) | | | | | Neg. (%) | Indif. (%) | Pos. (%) |
|---|---|---|---|---|---|---|---|---|---|
| | | 1 | 2 | 3 | 4 | 5 | | | |
| P1 | I would like to repeat this experience in other courses. | 3.09 | 6.79 | 11.1 | 32.1 | 46.3 | 9.88 | 11.11 | 78.4 |
| P2 | I considered the interdisciplinary relation positive to my learning. | 0.62 | 7.41 | 5.56 | 35.1 | 51.2 | 8.03 | 5.56 | 86.4 |
| P3 | The deadline for the project completion was satisfactory. | 4.32 | 13.5 | 24.0 | 33.3 | 23.4 | 17.9 | 24.07 | 56.7 |
| P4 | At the end of the project, I fulfilled the goals that I pursued. | 1.24 | 6.17 | 9.88 | 58.0 | 24.0 | 7.41 | 9.88 | 82.1 |
| P5 | I did not feel overwhelmed with the realization of PBL. | 12.3 | 29.0 | 34.5 | 16.0 | 7.41 | 41.3 | 34.57 | 23.4 |
| **Assess how the PBL impacted ...** | | | | | | | | | |
| C1 | ...in your ability to solve power electronics problems. | 2.47 | 4.94 | 19.1 | 51.2 | 18.5 | 7.41 | 19.14 | 69.7 |
| C2 | ...in your ability to solve control problems. | 3.70 | 3.70 | 14.8 | 43.8 | 33.9 | 7.41 | 14.82 | 77.7 |
| C3 | ...in your ability to make engineering decisions. | 1.85 | 2.47 | 16.0 | 50.6 | 29.0 | 4.32 | 16.05 | 79.6 |
| C4 | ...in your ability to seek information for yourself. | 0.62 | 0.62 | 14.2 | 48.1 | 36.4 | 1.24 | 14.20 | 84.5 |
| C5 | ...in your ability to solve problems presented in class. | 1.24 | 3.09 | 19.1 | 46.9 | 29.6 | 4.32 | 19.14 | 76.5 |
| **By participating in the group ...** | | | | | | | | | |
| G1 | ...I felt that cooperation helped to develop new ideas. | 1.85 | 5.56 | 12.3 | 43.2 | 36.4 | 7.41 | 12.35 | 79.6 |
| G2 | ...usually I recognize the skills of my colleagues. | 0.62 | 0 | 4.32 | 46.9 | 46.9 | 0.62 | 4.32 | 93.8 |
| G3 | ...I felt that everybody collaborated in the search for solutions. | 3.09 | 9.88 | 15.4 | 34.5 | 36.4 | 12.9 | 15.43 | 70.9 |
| G4 | ...I appreciate the union created between people. | 0 | 2.47 | 9.26 | 37.6 | 50.0 | 2.47 | 9.26 | 87.6 |
| G5 | ...I increase the esteem for my colleagues. | 0 | 2.47 | 12.3 | 41.3 | 43.2 | 2.47 | 12.35 | 84.5 |
| **In general...** | | | | | | | | | |
| E1 | ...I felt comfortable when I had to face unforeseen situations. | 1.85 | 15.4 | 26.5 | 42.5 | 13.5 | 17.2 | 26.54 | 56.1 |
| E2 | ...I managed to minimize the negative effects of adversity. | 1.85 | 6.79 | 20.3 | 54.9 | 16.0 | 8.64 | 20.37 | 70.9 |
| E3 | ...I kept me balanced facing stressful situations | 3.70 | 8.03 | 25.3 | 42.5 | 19.1 | 11.7 | 25.31 | 61.7 |
| E4 | ...I am aware of my intellectual abilities. | 0 | 1.85 | 11.7 | 55.5 | 29.6 | 1.85 | 11.73 | 85.1 |
| E5 | ...I believe I have skills to be successful. | 0.62 | 3.09 | 6.79 | 46.3 | 43.2 | 3.70 | 6.79 | 89.5 |
| **Therefore...** | | | | | | | | | |
| R1 | ...I have willpower to accomplish my goals. | 0 | 0.62 | 5.56 | 46.3 | 46.9 | 0.62 | 5.56 | 93.2 |
| R2 | ...I involved all my skills in the work that was done. | 1.85 | 1.85 | 17.9 | 50.0 | 27.7 | 3.70 | 17.90 | 77.7 |
| R3 | ...I feel accomplished as a student | 1.24 | 6.79 | 16.6 | 43.8 | 30.8 | 8.03 | 16.67 | 74.9 |
| R4 | ...I feel that instructors contributed to my development. | 1.24 | 4.32 | 9.88 | 38.8 | 45.0 | 5.56 | 9.88 | 83.9 |
| R5 | ...I feel that every year I improve my skills. | 0 | 1.24 | 6.79 | 38.8 | 52.4 | 1.24 | 6.79 | 91.3 |

It is considered that humanization exists at a level of abstraction beyond those that builds the individuality and the cooperation and serves, for the purposes of this work, as a way of grouping skills non-technical [9], [11].

After the assumption of such higher order construct, it is possible to infer the hypotheses regarding the constructs defined in the survey [13]. According to [12], the hypotheses of quantitative origin are predictions made by the researcher regarding the expected relationships between variables, and their confirmation depends on the statistical procedure employed by the researchers on the population of a study.

With the objective of understanding the formation of knowledge with the application of a PBL methodology, taking into account the humanization of the process, hypotheses to investigate these relationships are suggested. The hypotheses formulated and their rationale are presented below:

**Hypothesis A:** Humanization is a common factor of self-esteem;

This presupposition seeks to understand the question of the student's personal satisfaction with himself in the humanistic aspect of teaching [14]–[16].

It is assumed that individuality is an important aspect in the training of the engineer, responsible for helping or not his learning process, supported by aspects related to self-esteem, self-actualization and emotional background of the student.

**Hypothesis B:** Humanization is a common factor of self-realization;

This assumption examines the student's relationship with his or her tendency to develop their growth capacities [14], [17], [18]. This hypothesis has another aspect concerning the assumption of the importance of individuality in the formation of the engineer.

**Hypothesis C:** Humanization is a common factor of cooperation;

This assumption is based on the transversal competences of the 21st century, which value cooperation as an integral part of the modern world [11], [19].

In addition, it is also based on the ideas of [20] and [21], authors that address the fragmentation of the world's existing knowledge, and the importance of integration for the society progress.

**Hypothesis D:** Humanization has a positive influence on PBL;

Humanization as the foundation of the PBL is based on the concepts presented by [22], who cites the importance of an engineer involved with humanitarian and social aspects, who is integrally involved in the community in a manner that the knowledge he acquires is useful.

This hypothesis also seeks in [10] his confirmation, an author that addresses the importance understanding the student's own role on the word before the learning. This aspect was also shown in the Maslow's pyramid [23] with the assumption that knowledge will only be acquired if all human needs are satisfied. For [12] the complementation of the technical and human aspects is fundamental on the training of a student for the society.

**Hypothesis E:** PBL positively influences learning;

On the assumption that deals with the positive influence of learning in the PBL, it is possible to resort to all the authors that have already applied the methodology in the Electrical Engineering context [13], [24]–[32].

Relating to these hypotheses, it is suggested that the proposed and applied PBL has its roots in a humanist basis, formed by individuality (self-esteem and self-realization) and by cooperation among students, and thus sustained by this humanization, PBL serves as the basis for learning. As shown in Figure 1[1].

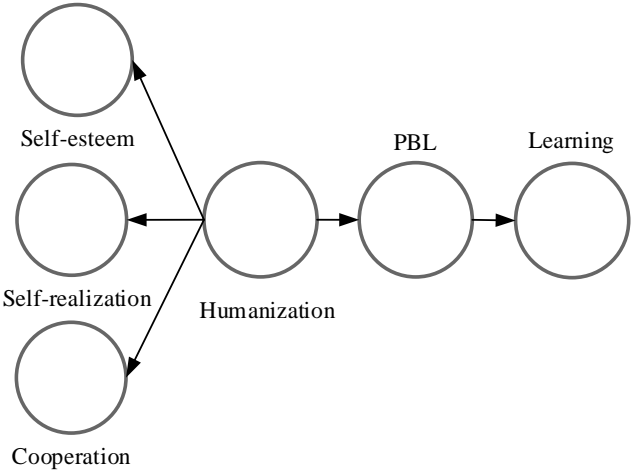

Fig. 1. Analyzed model based on the literature hypotheses.

### C. Structural Equation Modeling

The structural equation modeling is a multivariate analysis of second generation which seeks to understand the relationship between two or more variables simultaneously in order to assess the structural composition of the analyzed aspect [29], [30].

The PLS-PM algorithm (Partial Least Squares Path Modeling) was used in the modeling of structural equations. According to [31], this algorithm has proved to be adequate when the analysis has an exploratory aspect. In addition, current studies show that the technique has been consolidating increasingly in exploratory studies of social sciences [30].

The hypotheses have been analyzed using the PLS-PM algorithm presented in Algorithm 1, and are shown in Table I, defining the path coefficient (linear regression of the scores) between each connected latent variable.

ALGORITHM I
PLS-PM

PLS Path Modeling with path scheme, standardized latent variable scores and OLS regressions [32].

**Input:** $X = [X_1, ..., X_q, ..., X_Q]$, i.e. Q blocks of centered manifest variables;

**Output:** $w_q$, $\xi_q$, $\beta_j$ ;

1: **for all** q = 1, ... , Q do

2:    initialize $w_q$

3:    $v_q \propto \pm \sum_{p=1}^{P_q} w_{pq} x_{pq} = \pm X_q w_q$

4:    $e_{qq'} = \begin{cases} cor(v_q, v_{q'}) \\ (v_q \, '\, v_{q'})^{-1} v_q v_{q'} \end{cases}$

5:    $\vartheta_q \propto \sum_{q' \backslash 1}^{Q'} e_{qq'} v_{q'}$

6:    update $w_q$, $w_{pq} = \text{cov}(x_{pq}, \vartheta_q)$

7: **end for**

8: **Steps 1-7 are repeated until convergence** on the outer weights is achieved, i.e. until:

$$\max\{w_{pq, current\_iteration} - w_{pq, previous\_iteration}\} < \Delta$$

where $\Delta$ is a convergence tolerance usually set at 0.0001 or less

9: **Upon convergence:**

(1) for each block the standardized latent variable scores are computed as weighted aggregates of manifest variables:

$$\hat{\xi}_q \propto X_q w_{q'}$$

(2) for each endogenous latent variable $\xi_j (j = 1, ..., J)$ the vector of path coefficients is estimated by means of OLS regression as:

$$\beta_j = (\hat{\Xi}' \hat{\Xi})^{-1} \hat{\Xi}' \hat{\xi}_{j'}$$

where $\hat{\Xi}$ includes the scores of the latent variables that explain the *j*-th endogenous latent variable $\xi_j$, and $\hat{\xi}_j$ is the latent variable score of the *j*-th endogenous latent variable.

TABLE I
PROCESS ASSESSMENT SURVEY

| Hypothesis | Interaction | Influence [2] |
|---|---|---|
| $H_A$ | Humanization → Self-esteem | 0.874 |
| $H_B$ | Humanization → Self-realization | 0.866 |
| $H_C$ | Humanization → Cooperation | 0.753 |
| $H_E$ | Humanization – PBL | 0.733 |
| $H_D$ | PBL → Learning | 0.725 |

The results were validated using the methodology proposed by [33], evaluating the indicators reliability (indicators with loading bigger than 0,7), the internal consistence reliability (indicators of a same dimension share a high correlation) and the discriminant validity (indicators are better represented by the dimensions they were allocated). Also the result was tested with and Bias-Correct and Accelerated bootstrapping process [34].

After assessing the results generated by the PLS-PM it was possible to validate Humanization as a common factor of the individual (self-esteem relationship of 0.874 and self-realization relationship of 0.866) and social (cooperation of 0.753) aspects. Humanization also had a high influence on PBL (0.733). Finally, PBL also showed significance influence on learning (0.725).

---

[1] Arrows connecting humanization to self-esteem, self-realization and cooperation are in the opposite direct because they are considered all common effects of humanization.

[2] All values had a p-value < 0,001.

## IV. PLS AGENT (PBL CLASSROOM MODEL)

The agent-based simulation (ABM) is part of a class of computational models to simulate the actions and interactions of autonomous agents as a way of analyzing their role as a whole. In the proposition made by [7], PLS-PM can be used as a way of quantifying the cause-effect relationships between the studied phenomena and later be used as input parameters in ABM.

The results of the actions to which the agents are subject are given by probabilities. The probability of an event to occur with an agent is as a sum of the PLS-PM path coefficients ($\beta_{i,j}$) of the latent variable connected to that event divided by the maximum expected score of event, as shown in (1)

$$P_i = \frac{\sum_{j=1}^{J} \beta_{i,j}}{\max(\sum_{j=1}^{J} \beta_{i,j})} \quad (1)$$

Thus, the interpretation of the agent on the studied effect is based on the dimensions previously defined by the survey and quantized by the PLS-PM algorithm resulting path coefficients. In the proposed model, the main effect to be analyzed is the learning, considering different design points for the humanization and for the educational process (PBL), in a model called "PBL Classroom Model".

The path coefficients used in the process are the total effects, defined even if the dimensions are not directly connected, i.e.: even if humanization is not directly connected on learning, it has an indirect influence effect calculated as (2), where the arrows points the direction of the influence, so the total effect of humanization on learning is 0,532.

$$(Humanization \rightarrow Learning) =$$
$$(Humanization \rightarrow PBL).(PBL \rightarrow \text{Learning}) \quad (2)$$

The probability of humanization considers the influences of self-esteem, self-realization and cooperation. For each of the causes of humanization is assigned a maximum score (in this case 10 multiplied by [34]) which is multiplied by the total effect of each latent variable, the same is done for learning, which depends on humanization and PBL. This way the $\max(\sum_{j=1}^{J} LVscore_{i,j})$ is shown on Table II.

TABLE II
AGENT MAXIMUM MODEL

| Dimension | Maximum Score | Total Effects | Results | Humanization Max Value |
|---|---|---|---|---|
| Cooperation | 10 | 0.753 | 7550 | 24950 |
| Self-esteem | 10 | 0.874 | 8740 | |
| Self-realization | 10 | 0.866 | 8660 | |
| | | | | **Learning Max Value** |
| Humanization | 10 | 0.532 | 5320 | 12570 |
| PBL | 10 | 0.725 | 7250 | |

Having the maximum scores calculated, probability of humanization and learning are calculated as shown in (3) and (4). Note that the probability of learning depends on the probability of humanization.

$$P_{humanization} = \frac{\begin{array}{c}0.874.(self\text{-}esteem) + 0.866.(self\text{-}realization) \\ + 0.753.(cooperation)\end{array}}{24950} \quad (3)$$

$$P_{learning} = \frac{0.532.(humanization) + 0.725.(PBL)}{12570} \quad (4)$$

The simulation parameters are presented in Table III, with the operator having to previously define the experimental factors values and the control variables to obtain the desired response variables. In the objective of this work, it is desired to discover the diffusion factor of learning for different values of humanization and quality of the applied methodology.

Following [7], [35] principles, the simulation investigation should follow the 3k-factorial design. This design requires that each experimental factor has one low, one medium, and one high value. The simulation experiments perform each possible combination based on these parameters. Both humanization and PBL attributes have factor values of 2, 5, and 8. Existing parameters, their scales and experimental design values are shown in Table III.

TABLE III
EXPERIMENTAL DESIGN PARAMETERS

| Parameters | Scales | Experimental Design |
|---|---|---|
| **Experimental Factors** | | |
| Cooperation | ∈ [0,...,10] | (2, 5, 8) |
| Self-esteem | ∈ [0,...,10] | (2, 5, 8) |
| Self-realization | ∈ [0,...,10] | (2, 5, 8) |
| PBL | ∈ [0,...,10] | (2, 5, 8) |
| **Control variables** | | |
| Agents number | N | 961 |
| Link-chance | ∈ [0,...,1] | random ∈ [0.3,...,0.7] |
| **Response Variable** | | |
| Diffusion rate | ∈ [0,...,1] | |

The simulation in based on a single grid network model, where every grid position is occupied with and immobile agent. Every agent is allowed to link with a maximum of 4 agents in its adjacent cells (Von Neumann topology) considering the link-chance probability, a probability that two neighbors have of connecting.

Only one kind of agent is considered on the model, the student agent. At every step of the simulation the student is able susceptible to two events: humanization and learning.

During the simulation the following consideration was made: agents with high index of cooperation when connecting with agents also high index of cooperation, can share knowledge, generating a possibility of activation of the learning event. The initial possibility of the agents to connect is given by the link-chance, considered random during the simulations ranging from 30% to 70%.

The following rules govern the simulation:

**Agents:** students
**Attributes:** learned; humanized
**On creation:** random chance of creating bonds to its classroom colleagues (neighbors)
**Step:** At each step of the simulation the students decide whether or not to change their attributes:

    • The student can learn if he has enough possibility and have not tried to learn before on his own;
    • The student can be humanized if he has enough possibility of humanization and have not tried to be humanized before;
    • The student can connect with its neighbors if both have succeeded in the link-chance possibility. After the connection, if both students have enough level of cooperation, they can transfer knowledge.

The simulation scenario is based on an active learning environment based on the previous defined Project Based Learning Agile.

### A. Simulation Results

The simulation was conducted using the Mesa Python library [36] and following the steps described in [35] as validation of the model.

The model is illustrated at the Figure 2 to a medium design point scenario (factors values of 5), where the squares represents the students who have not suffered an event, small circles are students who only learned (blue) or only humanized (red), and big circles are students who got both learned and humanized.

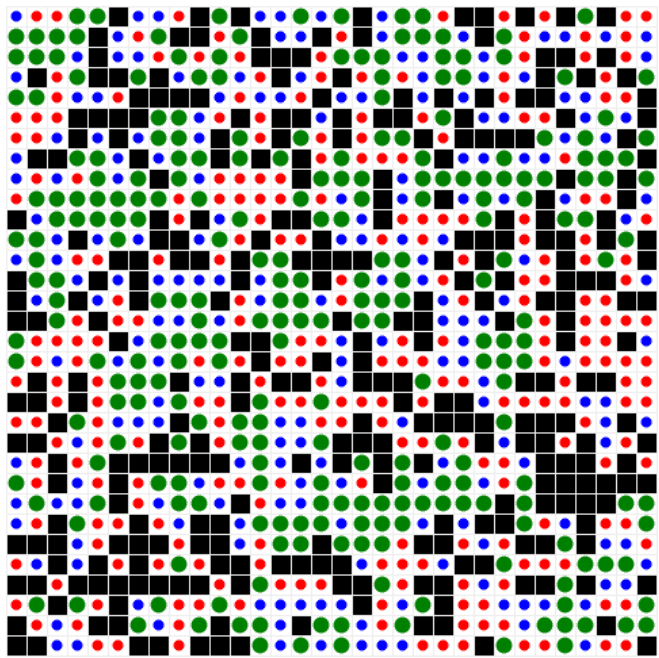

Fig. 2. Model simulation for a medium design point scenario.

First an error variance analysis must be conducted in order to verify the number of necessary runs per model setting. As proposed by [35] the variation overs over increased number of runs must be analyzed. For this analyses 3 settings were used: a low design point (cooperation = 2; self-esteem = 2; self-realization = 2; PBL = 2. link-chance = 30%), a middle design point (cooperation = 5; self-esteem = 5; self-realization = 5; PBL = 5; link chance = 50%) and a high design point (cooperation = 8, self-esteem = 8, self-realization = 8, PBL = 8, link-chance = 70%). The results are shown in Table IV.

The low standard deviation (SD) and coefficient of variation (CV) change from 300 to 500 runs for all design points shows that the model tends towards the stabilization and that it is adequate to proceed with the simulations using 300 runs. The defined number of runs is used on the posterior simulations of this paper.

TABLE IV
AGENT MAXIMUM MODEL

| Design point | | Diffusion rate | | | | |
|---|---|---|---|---|---|---|
| | | 1 run | 50 runs | 100 runs | 300 runs | 500 runs |
| Low | Mean | 0.0 | 0.10 | 0.11 | 0.13 | 0.13 |
| | SD | - | 0.04 | 0.03 | 0.02 | 0.02 |
| | CV | - | 0.37 | 0.26 | 0.19 | 0.18 |
| Medium | Mean | 0.28 | 0.44 | 0.47 | 0.51 | 0.51 |
| | SD | - | 0.11 | 0.13 | 0.10 | 0.10 |
| | CV | - | 0.26 | 0.27 | 0.21 | 0.21 |
| High | Mean | 0.64 | 0.82 | 0.87 | 0.91 | 0.92 |
| | SD | - | 0.26 | 0.21 | 0.17 | 0.16 |
| | CV | - | 0.31 | 0.24 | 0.18 | 0.17 |

Having decided the number of necessary runs, Table V contain eta squared effect sizes $\eta^2$ of each experimental factor on learning diffusion rate. Results shown that PBL has the higher impact on the diffusion rate ($\eta^2 = 0.61$), while cooperation, self-esteem and self-realization, all being part of a high order construct (humanization) have lower individual impacts.

However, it is important to notice that cooperation has an small advantage on its effect size ($\eta^2 = 0.12$) when compared to the individuals aspects ($\eta^2 = 0.08$).

TABLE V
ETA SQUARED ($\eta^2$) EFFECT SIZE

| Dimension | Cooperation |
|---|---|
| Cooperation | 0.12 |
| Self-esteem | 0.08 |
| Self-realization | 0.08 |
| PBL | 0.61 |

According to [35] a sensitivity analysis is an important step of the simulation to understand the agents interactions. Figure 3 shows the variation of the diffusion rate of the knowledge for different levels of link-chance, which is the same as raising the possibility to student's interchange knowledge when the humanization level is high enough.

Figure 4 shows the violin plot of learning diffusion rate for different link-chance levels (30% to 40%; 40% to 50%, 50% to 60). From the link-chance sensitive analysis it is possible to verify that the maximum rate of learning diffusion only occurs when students share acquired knowledge with each other.

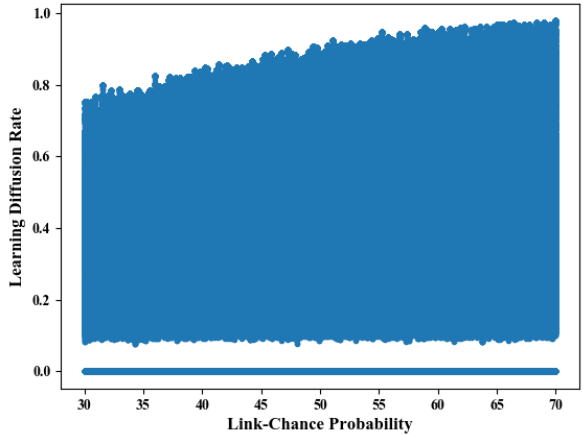

Fig. 3. Learning diffusion rate variation between 30% and 70% of link-chance.

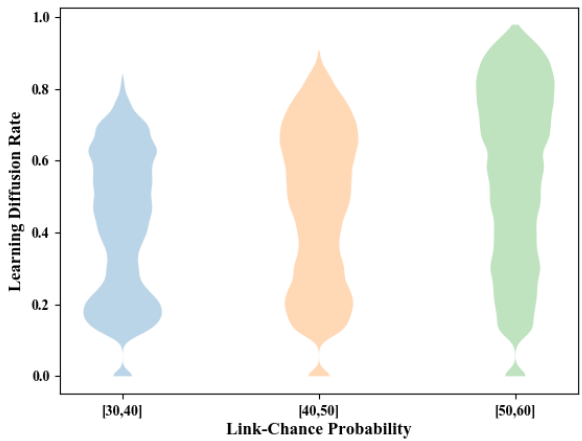

Fig. 4. Learning diffusion rate for different link-chance levels

After, two more simulations were conducted. One regarding the learning diffusion rate for different humanization levels, allow the PBL levels to vary (2, 5, 8). The results are shown in Figure 5, and show that the humanization is directly connected to the learning.

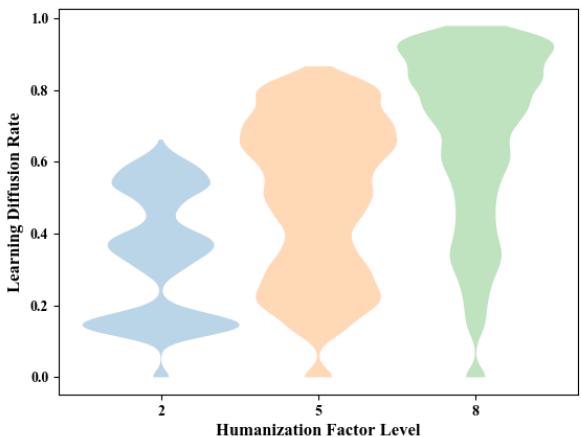

Fig. 5. Learning diffusion rate for different humanizations levels.

The second one regarding the learning diffusion rate for different methodology quality levels (PBL), allows the humanization to vary (2, 5, 8). The results are shown in Figure 6, it is possible to note that even with a high methodology quality level, the diffusion rate only reaches its maximum for a small sample (when the humanization is at 8).

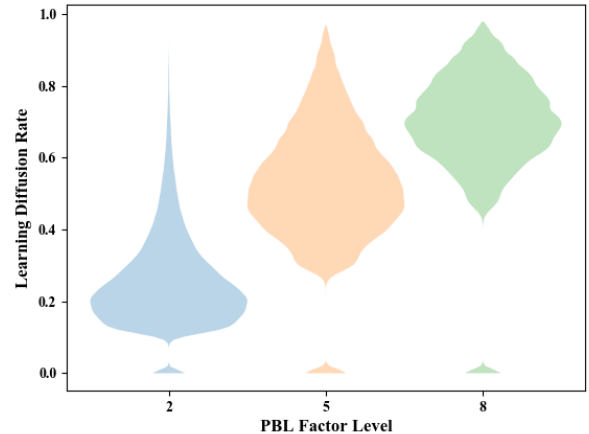

Fig. 6. Learning diffusion rate for different methodology quality levels (PBL).

## IV. CONCLUSION

This paper proposed and evaluated an agent-based simulation of the learning dissemination on a Project-Based Learning context considering the human aspects. To raise the input parameters of the simulation, an active learning called PBL[A] (Project-Based Learning Agile) was applied on the Regional University of Blumenau (FURB) during five consecutive semesters and the students were invited to answer a survey on the end of the application.

The survey answers were first submitted to a second generation multivariate analysis technique known of PLS-PM (partial least squares path modeling), which aims to identify relationships between non-observed variables. The results in the form of causal relationships were then used as input parameters to the agent-based simulation, in a model called "PBL Classroom Model".

In the simulation process, students with a high index of cooperation were allowed to exchange knowledge with others students with also a high index of cooperation, increasing the learning diffusion rate.

From the simulation results it was possible to verify that the learning diffusion rates are directly related to the humanization (self-esteem, self-realization and cooperation) and to the methodology quality. It was also possible to verify that the diffusion rates achieve higher values when the cooperation between students is higher, suggesting that even with a high quality teaching methodology, it's necessary that the students share knowledge between them to achieve the maximum learning.

The results presented in this study suggest that humanization, and mainly cooperation, is an important part in the educational formation process of a new electrical engineer, and that to achieve a higher learning diffusion rate, the teaching-learning relationship is better when also takes into account the human nature.

## References

[1] C. M. Vest, "Context and Challenge for Twenty-First Century Engineering Education," *J. Eng. Educ.*, vol. 97, no. 3, pp. 235–236, Jul. 2008.

[2] J. Taylor, "Engineering the information age," *IEE Rev.*, vol. 44, no. 6, pp. 250–252, Nov. 1998.

[3] M. Castells, "The Rise of the Network Society," *Massachusetts Blackwell Publ.*, vol. I, p. 594, 2010.

[4] S. M. Malcom, "The human face of engineering," *J. Eng. Educ.*, vol. 97, no. July, pp. 237–238, Jul. 2008.

[5] J. W. Pellegrino, "Rethinking and Redesigning Curriculum, Instruction and Assessment: What Contemporary Research and Theory Suggests," *A Pap. Comm. by Natl. Cent. Educ. Econ. New Comm. Ski. Am. Work.*, no. November, pp. 1–15, 2006.

[6] E. F. Redish and K. A. Smith, "Looking Beyond Content: Skill Development for Engineers," *J. Eng. Educ.*, vol. 97, no. 3, pp. 295–307, Jul. 2008.

[7] S. Schubring, I. Lorscheid, M. Meyer, and C. M. Ringle, "The PLS agent: Predictive modeling with PLS-SEM and agent-based simulation," *J. Bus. Res.*, vol. 69, no. 10, pp. 4604–4612, Oct. 2016.

[8] M. Fowler and J. Highsmith, "The agile manifesto," *Softw. Dev.*, 2001.

[9] J. J. Duderstadt, "Engineering for a Changing World," in *Holistic Engineering Education*, New York, NY: Springer New York, 2010, pp. 17–35.

[10] E. Morin, "Seven Complex Lessons in Education for the Future: Education on the Move," p. 92, 2001.

[11] S. D. Sheppard, J. W. Pellegrino, and B. M. Olds, "On Becoming a 21st Century Engineer," *J. Eng. Educ.*, vol. 97, no. 3, pp. 231–234, Jul. 2008.

[12] J. B. Toro, *Códigos da modernidade: capacidades e competências mínimas para participação produtiva no século XXI*. Porto Alegre: Fundação Maurício Sirotsky Sobrinho, 1988.

[13] L. O. Seman, G. Gomes, and R. Hausmann, "Statistical Analysis Using PLS of a Project-Based Learning Application in Electrical Engineering," *IEEE Lat. Am. Trans.*, vol. 14, no. 2, pp. 646–651, Feb. 2016.

[14] E. Morin, "Os setes saberes necessários à educação do futuro," 2001.

[15] D. Kahneman, "Rápido e devagar: duas formas de pensar," 2012.

[16] U. Neisser *et al.*, "Intelligence: Knowns and unknowns.," *Am. Psychol.*, vol. 51, no. 2, pp. 77–101, 1996.

[17] A. H. Maslow, "Introdução à psicologia do ser," 1968.

[18] R. Ryan and E. Deci, "Intrinsic and extrinsic motivations: Classic definitions and new directions," *Contemp. Educ. Psychol.*, 2000.

[19] J. W. Pellegrino and M. L. Hilton, *Education for Life and Work: Developing Transferable Knowledge and Skills in the 21st Century*. Washington, D.C.: National Academies Press, 2012.

[20] M. Ridley, "The rational optimist." 2010.

[21] F. A. Hayek, "The use of knowledge in society.," *Am. Econ. Rev.*, no. 35, 1945.

[22] W. A. Bazzo, "Uma nova equação civilizatória x Problemas contemporâneos da educação. Mensagem do coordenador," 2016. [Online]. Available: http://www.nepet.ufsc.br/. [Accessed: 13-Apr-2016].

[23] A. Maslow, *Toward a psychology of being*. Princeton: D Van Nostrand, 1962.

[24] J. Kim, "An Ill-Structured PBL-Based Microprocessor Course Without Formal Laboratory," *IEEE Trans. Educ.*, vol. 55, no. 1, pp. 145–153, Feb. 2012.

[25] N. Hosseinzadeh and M. R. Hesamzadeh, "Application of Project-Based Learning (PBL) to the Teaching of Electrical Power Systems Engineering," *IEEE Trans. Educ.*, vol. 55, no. 4, pp. 495–501, Nov. 2012.

[26] J. W. Thomas, "A review of research on project-based learning," 2000.

[27] R. H. Chu, D. D.-C. Lu, and S. Sathiakumar, "Project-Based Lab Teaching for Power Electronics and Drives," *IEEE Trans. Educ.*, vol. 51, no. 1, pp. 108–113, 2008.

[28] W. Powell, P. Powell, and W. Weenk, "Project Led Engineering Education," 2003.

[29] D. G. Lamar, P. F. Miaja, M. Arias, A. Rodriguez, M. Rodriguez, and J. Sebastian, "A project-based learning approach to teaching power electronics: Difficulties in the application of Project-Based Learning in a subject of Switching-Mode Power Supplies," in *IEEE EDUCON 2010 Conference*, 2010, pp. 717–722.

[30] L. E. M. Brackenbury, L. A. Plana, and J. Pepper, "System-on-Chip Design and Implementation," *IEEE Trans. Educ.*, vol. 53, no. 2, pp. 272–281, May 2010.

[31] E. Guzman Ramirez, I. Garcia, E. Guerrero, and C. Pacheco, "A tool for supporting the design of DC-DC converters through FPGA-based experiments," *IEEE Lat. Am. Trans.*, vol. 14, no. 1, pp. 289–296, Jan. 2016.

[32] A. Kumar, S. Fernando, and R. C. Panicker, "Project-Based Learning in Embedded Systems Education Using an FPGA Platform," *IEEE Trans. Educ.*, vol. 56, no. 4, pp. 407–415, Nov. 2013.

[33] J. F. Hair, M. Sarstedt, C. M. Ringle, and J. A. Mena, "An assessment of the use of partial least squares structural equation modeling in marketing research," *J. Acad. Mark. Sci.*, vol. 40, no. 3, pp. 414–433, May 2012.

[34] B. Efron and R. J. Tibshirani, "An Introduction to the Bootstrap," 1993.

[35] I. Lorscheid, B.-O. Heine, and M. Meyer, "Opening the 'black box' of simulations: increased transparency and effective communication through the systematic design of experiments," *Comput. Math. Organ. Theory*, vol. 18, no. 1, pp. 22–62, Mar. 2012.

[36] D. Masad and J. Kazil, "MESA: An Agent-Based Modeling Framework," *Proc. 14th Python Sci. Conf. (SCIPY 2015)*, no. Scipy, pp. 53–60, 2015.

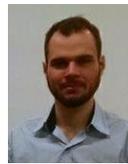

**Laio Oriel Seman** received the master degree in Electrical Engineering from the Regional University of Blumenau in 2013. He is currently a PhD candidate at the Federal University of Santa Catarina. His interests include multivariate data analysis and structural equation modeling.

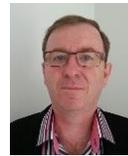

**Romeu Hausmann** received the master and PhD degree in Electrical Engineering from the Federal University of Santa Catarina in 2000 and 2011 respectively. He is currently a professor at the Regional University of Blumenau (FURB), focusing his researches in DC-DC and multilevel converters.

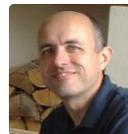

**Eduardo Augusto Bezerra** PhD in Computer Engineering, University of Sussex, England, UK. He is a lecturer at UFSC, Brazil, and visiting researcher at LIRMM, France. His research interests include embedded systems for space applications, computer architecture, and reconfigurable systems (FPGAs).